\documentclass[12pt,draftcls,journal,letterpaper,onecolumn]{IEEEtran}

\usepackage{amsfonts,cite,graphicx}

\usepackage[cmex10]{amsmath}
\newlength{\plotwidth}
\DeclareMathOperator*{\argmin}{arg\,min}

\begin{document}
\title{Minimizing Sum-MSE Implies Identical Downlink and Dual Uplink Power Allocations}
\author{
Adam~J.~Tenenbaum,~\IEEEmembership{Student~Member,~IEEE,}
and~Raviraj~S.~Adve,~\IEEEmembership{Senior~Member,~IEEE}
\thanks{The authors are with the Edward S.~Rogers Sr.~Department of Electrical and
Computer Engineering, University of Toronto, 10 King's College Road,
Toronto, Ontario, Canada M5S 3G4. (e-mail:
\{adam,rsadve\}@comm.utoronto.ca).}
\thanks{This work has been submitted to the IEEE for possible publication. Copyright may be transferred without notice, after which this version may no longer be accessible.}
}
\maketitle

\begin{abstract}
In the multiuser downlink, power allocation for linear precoders that
minimize the sum of mean squared errors under a sum power constraint is
a non-convex problem. Many existing algorithms solve an equivalent
convex problem in the virtual uplink and apply a transformation based
on uplink-downlink duality to find a downlink solution. In this letter,
we analyze the optimality criteria for the power allocation subproblem
in the virtual uplink, and demonstrate that the optimal solution leads
to identical power allocations in the downlink and virtual uplink. We
thus extend the known duality results and, importantly, simplify the
existing algorithms used for iterative transceiver design.
\end{abstract}


\section{Introduction}
\IEEEPARstart{I}n the multiuser multiple-input, multiple-output (MIMO)
downlink, linear transmit/receive processing to minimize the sum of
mean squared errors (sum-MSE) under a sum power constraint is a
well-studied problem. When formulated as a precoder design problem,
with implicit minimum-MSE (MMSE) receive matrices, the sum-MSE is a
non-convex function of the downlink precoders.  A standard approach
used to solve the problem is to find an equivalent formulation in the
\textit{virtual uplink}, wherein the roles of transmitter and receiver
are exchanged~\cite{SSJB05,KTA06,MJHU06,SSB07} .  In the virtual
uplink, the receiver is the Wiener filter and the power allocation
subproblem is convex.

The equivalence of the downlink and virtual uplink problems are enabled
by an \textit{uplink-downlink duality} result for the MSE of each data
stream.  Duality results for linear precoding systems were first
presented for signal-to-interference-plus-noise ratios (SINR)
in~\cite{SB04} with single-antenna receivers.  This work was later
extended to MSEs and systems with multiple receive antennas
in~\cite{SSJB05,KTA06} and subsequently generalized in~\cite{HJU09}.
The algorithms we focus on are based on iterating between the downlink and
virtual uplink. Crucially, a common feature is a transformation of the
resulting power allocation in the virtual uplink to the downlink, while
achieving the same MSE in each stream in both systems. This
transformation requires solving a matrix equation in each iteration.

In this letter, we use the Karush-Kuhn-Tucker (KKT) conditions for the
power allocation subproblem in the virtual uplink to show that at the
optimal point, the powers allocated to each data stream in both the
downlink and virtual uplink are identical.  This result extends the
known dualities in the multiuser MIMO case. Importantly, this also
eliminates the need for the uplink-to-downlink power transformation,
significantly simplifying the algorithms used.

This paper is organized as follows: Section~\ref{section:bg} describes
the system model and existing algorithms for minimizing the sum-MSE
using uplink-downlink duality. In Section~\ref{section:proof}, we
present the KKT conditions for the virtual uplink power allocation
subproblem, and use the resulting expressions to prove the equality of
the downlink and virtual uplink power allocations.
Section~\ref{section:conclusions} wraps up the paper with some
conclusions.

\section{Background\label{section:bg}}
\subsection{System Model with Linear Precoding}
\begin{figure}
\centering
\includegraphics[width=\plotwidth]{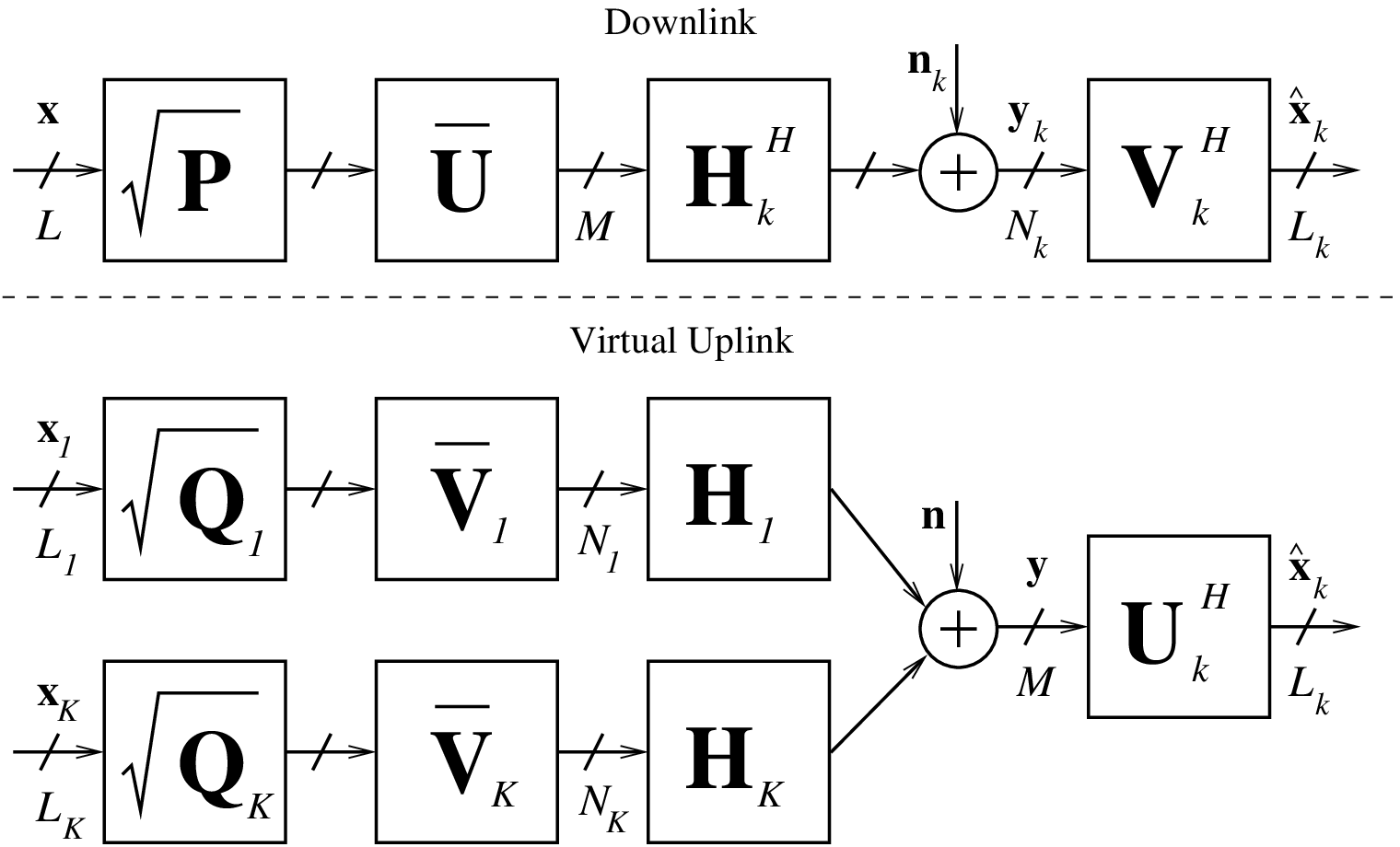}
\caption{Processing for user $k$ in downlink and virtual uplink.}\label{fig:systemmodel}
\end{figure}

In the linear precoding system, illustrated in
Fig.~\ref{fig:systemmodel}, a base station with $M$ antennas transmits
to $K$ decentralized mobile users over flat wireless channels; user $k$
has $N_k$ receive antennas.
The channel between the transmitter and user $k$ is represented by the
$N_{k}\times M$ matrix $\mathbf{H}_{k}^{H}$, and the overall $N\times
M$ channel matrix is $ \mathbf{H}^H$, with
$\mathbf{H}=\left[\mathbf{H}_{1},\dots,\mathbf{H}_{K} \right]$.  User
$k$ receives $L_k$ data symbols $\mathbf{x}_{k}=\left[x_{k1}, \dots,
x_{kL_{k}}\right]^T$ from the base station, and the vector $\mathbf{x}
= \left[\mathbf{x}_1^T, \dots, \mathbf{x}_K^T\right]^T$ comprises
independent symbols with unit average energy
($\mathbb{E}\left[\mathbf{x}\mathbf{x}^{H}\right]=\mathbf{I}_{L}$ (the
$L\times L$ identity matrix), where $L=\sum_{k=1}^K L_k$).  User $k$'s
data streams are precoded by the $M\times L_k$ transmit filter
$\bar{\mathbf{U}}_{k} = \left[\bar{\mathbf{u}}_{k1}, \dots,
\bar{\mathbf{u}}_{kL_k}\right]$, where $\bar{\mathbf{u}}_{kj}$ is the
precoding beamformer for stream $j$ of user $k$ with
$\|\bar{\mathbf{u}}_{kj}\| = 1$.  These individual precoders are
combined in the $M\times L$ global transmitter precoder matrix
$\bar{\mathbf{U}} = \left[\bar{\mathbf{U}}_{1}, \dots,
\bar{\mathbf{U}}_{K}\right]$. Power is allocated to user $k$'s data
streams in the vector $\mathbf{p}_{k}=\left[p_{k1}, \dots, p_{kL_{k}}
\right]^T$ and $\mathbf{P}_k=\mathrm{diag}\left[\mathbf{p}_k\right]$;
we define the diagonal downlink power allocation matrix as
$\mathbf{P}=\mathrm{diag}\{\left[\mathbf{p}_1^T, \dots,
\mathbf{p}_K^T\right]\}$.  Based on this model, user $k$ receives a
length-$N_{k}$ vector
$\mathbf{y}_{k}^{DL}=\mathbf{H}_{k}^{H}\bar{\mathbf{U}}\sqrt{\mathbf{P}}\mathbf{x}
+ \mathbf{n}_{k}$, where the superscript $^{DL}$ indicates the
downlink, and $\mathbf{n}_k \sim
\mathcal{CN}(0,\sigma^2\mathbf{I}_{N_k})$ consists of zero-mean white
Gaussian noise.  To estimate its $L_{k}$ symbols $\mathbf{x}_{k}$, user
$k$ applies the $L_{k}\times N_{k}$ receive filter
$\mathbf{V}_{k}^{H}$, yielding the estimated symbols
$\hat{\mathbf{x}}_{k}^{DL}=\mathbf{V}_{k}^{H}\mathbf{H}_{k}^{H}\bar{\mathbf{U}}\sqrt{\mathbf{P}}
\mathbf{x}+\mathbf{V}_{k}^{H}\mathbf{n}_{k}$.

To minimize the sum-MSE in the multiuser MIMO downlink, we use the
virtual uplink, also illustrated in Fig.~\ref{fig:systemmodel}, where
each matrix is replaced by its conjugate transpose.  In this
transformed system, we imagine transmissions from mobile user $k$ that
propagate via the \textit{transpose channel} $\mathbf{H}_k$ to the base station.  The
transmit and receive filters for user $k$ become $\bar{\mathbf{V}}_{k}$
and $\mathbf{U}_{k}^{H}$ respectively, with normalized precoding
beamformers; i.e., $\|\bar{\mathbf{v}}_{kj}\| = 1$. Power is allocated
to user $k$'s data streams as $\mathbf{q}_{k}=\left[q_{k1}, \dots,
q_{kL_{k}}\right]^T$, with
$\mathbf{Q}_k=\mathrm{diag}\left[\mathbf{q}_k\right]$ and
$\mathbf{Q}=\mathrm{diag}\{\left[\mathbf{q}_1^T, \dots,
\mathbf{q}_K^T\right]\}$.  The received symbol vector at the base
station and the estimated symbol vector for user $k$ are
$\mathbf{y}^{UL} =
\sum_{i=1}^{K}\mathbf{H}_{i}\bar{\mathbf{V}}_{i}\sqrt{\mathbf{Q}_{i}}\mathbf{x}_{i}+\mathbf{n}$
and ${\mathbf{x}}_{k}^{UL} =
\sum_{i=1}^{K}\mathbf{U}_{k}^{H}\mathbf{H}_{i}\bar{\mathbf{V}}_{i}\sqrt{\mathbf{Q}_{i}}
\mathbf{x}_{i}+\mathbf{U}_{k}^{H}\mathbf{n}$,
respectively, with zero-mean white Gaussian noise $\mathbf{n} \sim
\mathcal{CN}(0,\sigma^2\mathbf{I}_{M})$.

\subsection{Minimum Sum-MSE Multiuser MIMO Linear Precoding}
\subsubsection{Convex Minimum Sum-MSE Precoder Design} 
The MSE matrix for user $k$ in the downlink using arbitrary precoder
and decoder matrices can be written as
\begin{equation}\label{eqn:msedlmodel}
\begin{split}
\mathbf{E}^{DL}_k & = \mathbb{E}\left[\left(\hat{\mathbf{x}}_k^{DL} -
\mathbf{x}_k\right)\left(\hat{\mathbf{x}}_k ^{DL}- \mathbf{x}_k\right)^H\right] \\
& = \mathbf{V}_{k}^{H}\mathbf{J}_k\mathbf{V}_{k} -\mathbf{V}_{k}^{H}\mathbf{H}_{k}^{H}
\tilde{\mathbf{U}}_k  -\tilde{\mathbf{U}}_k^{H}\mathbf{H}_{k}\mathbf{V}_{k} + \mathbf{I}_{L_k},
\end{split}
\end{equation}
where
$\mathbf{J}_k=\mathbf{H}_{k}^{H}\bar{\mathbf{U}}\mathbf{P}\bar{\mathbf{U}}^{H}
\mathbf{H}_{k}+\sigma^2\mathbf{I}$,
$\tilde{\mathbf{U}}_k=\bar{\mathbf{U}}_k\sqrt{\mathbf{P}_k}$, and data
and noise terms are assumed to be independent.  The individual MSE
terms are minimized using the MMSE receiver ${\mathbf{V}_{k}^{\star}}^H
= \tilde{\mathbf{U}}_k^H\mathbf{H}_k\mathbf{J}_k^{-1}$.  The resulting
MMSE matrix is ${\mathbf{E}^\star_k}^{DL} = \mathbf{I}_{L_k} -
\tilde{\mathbf{U}}_k^H\mathbf{H}_k\mathbf{J}_k^{-1}\mathbf{H}_k^H\tilde{\mathbf{U}}_k$,
and the minimum sum-MSE for any choice of $\tilde{\mathbf{U}}_k$ is
$\mathrm{SMSE}^{DL} = \sum_{k=1}^K
\mathrm{tr}\left[{\mathbf{E}^\star_k}^{DL}\right]$.

The problem of finding the sum-MSE minimizing precoders and power
allocations in the downlink under a sum power constraint
$\mathrm{tr}\left[\mathbf{P}\right] \le P_{\max}$ is non-convex when
MMSE receivers $\mathbf{V}_k^\star$ are defined as a function of
$\tilde{\mathbf{U}}$ due to the cross-coupling introduced by the
presence of all $\tilde{\mathbf{U}}_i$ terms in every $\mathbf{J}_k$.
The authors of~\cite{HJU09} demonstrate hidden convexity in the
downlink sum-MSE minimization problem by optimizing receive matrices
(using closed form MMSE precoders) and applying a modified cost
function.  However, this problem is more commonly solved via
transformation to the virtual uplink, which gives rise to several
equivalent problems that can be solved using convex optimization.  The
set of virtual uplink minimum sum-MSE precoders and power allocations
$\left\{\left(\bar{\mathbf{V}}_k,\mathbf{q}_k\right),k=1,\dots,K\right\}$
can be found jointly, by finding the optimum covariance matrices
$\mathbf{R}_k=\bar{\mathbf{V}}_k^H\mathbf{Q}_k\bar{\mathbf{V}}_k$ and
applying Cholesky or eigen-decomposition~\cite{SSB07}.  An alternative
approach finds the optimum precoders $\bar{\mathbf{V}}_k$ and power
allocations $\mathbf{q}_k$ in an iterative manner~\cite{KTA06,SSB07}.
The convexity of these problems originates from the decoupling of users
in the virtual uplink.  The MMSE matrix ${\mathbf{E}^\star_k}^{UL}$ for
user $k$ is found using the MMSE receiver
${\mathbf{U}^\star_k}^H=\tilde{\mathbf{V}}_k^H\mathbf{H}_k^H\mathbf{J}^{-1}$,
\begin{equation}\label{eqn:mmsevul} {\mathbf{E}^\star_k}^{UL} =
\mathbf{I}_{L_k} -
\tilde{\mathbf{V}}_k^H\mathbf{H}_k^H\mathbf{J}^{-1}\mathbf{H}_k\tilde{\mathbf{V}}_k,
\end{equation}
with $\tilde{\mathbf{V}}_k=\bar{\mathbf{V}}_k\sqrt{\mathbf{Q}_k}$ and
$\mathbf{J}=\sum_{k=1}^K\mathbf{H}_{k}\tilde{\mathbf{V}}_k\tilde{\mathbf{V}}_k^H\mathbf{H}_k^H
+ \sigma^2\mathbf{I}_M$.  The resulting minimum sum-MSE is
\begin{equation}
\begin{split}
\mathrm{SMSE}^{UL} & = \sum_{k=1}^K L_k - \mathrm{tr}\left[\mathbf{J}^{-1}
\sum_{k=1}^K \mathbf{H}_k\tilde{\mathbf{V}}_k\tilde{\mathbf{V}}_k^H\mathbf{H}_k^H\right]\\
& = L-M+\sigma^2\mathrm{tr}\left[\mathbf{J}^{-1}\right],
\end{split}
\end{equation}
which follows from
$\mathrm{tr}\left[\mathbf{AB}\right]=\mathrm{tr}\left[\mathbf{BA}\right]$,
linearity of the trace operator, and the definition of $\mathbf{J}$.
Minimizing the sum-MSE thus only requires minimization of
$\mathrm{tr}\left[\mathbf{J}^{-1}\right]$, which is convex for both the
power allocation subproblem in $\mathbf{q}_k$ and the joint precoder
design problem in covariance matrices $\mathbf{R}_k$.  In this letter,
we consider the former optimization problem, which is formally stated
as
\begin{equation}\label{eqn:optPowerAlloc}
\begin{split}
\left(q^\star_1,\dots,q^\star_L\right) = & \argmin_{q_1,\dots,q_L}
\mathrm{tr}\left[\left(\sum_{l=1}^L q_l\tilde{\mathbf{h}}_l\tilde{\mathbf{h}}_l^H +
\sigma^2\mathbf{I}_M\right)^{-1}\right]\\
\mathrm{s.t.} \quad & q_l \ge 0 \quad l=1,\dots,L, \;\; \sum_{l=1}^L q_l \le P_{\mathrm{max}},
\end{split}
\end{equation}
where we have defined the effective channel
$\tilde{\mathbf{H}}=\left[\mathbf{H}_1\bar{\mathbf{V}}_1, \dots,
\mathbf{H}_K\bar{\mathbf{V}}_K\right]=\left[\mathbf{\tilde{h}}_1,\dots
,\mathbf{\tilde{h}}_{L}\right]$.  Note that the columns in
$\tilde{\mathbf{H}}$ refer to the effective channel vectors for each
individual data stream $l=1,\dots,L$.

\subsubsection{Uplink-Downlink Duality}
The duality results in~\cite{KTA06,SSB07} show that any set of MSEs
that are achievable in the virtual uplink are also achievable in the
downlink under the same sum power constraint.  This duality can be
satisfied by factoring the downlink and virtual uplink receive
beamforming vectors for each data stream $l$ as $\mathbf{v}_l^H =
p_l^{-\frac{1}{2}}\beta_l\bar{\mathbf{v}}_l^H$ and $\mathbf{u}_l^H =
q_l^{-\frac{1}{2}}\beta_l\bar{\mathbf{u}}_l^H$, where $\beta_l > 0$.
Any set of feasible MSEs $\mathbf{\varepsilon} =
\mathrm{diag}\left\{\varepsilon_1,\ldots,\varepsilon_L\right\}$ can
then be achieved by the downlink and virtual uplink power allocations
\begin{equation}\label{eqn:poweralloc}
\begin{split}
\mathbf{p}&=\sigma^{2}(\mathbf{\varepsilon} - \mathbf{D} -\mathbf{\beta}^2\mathbf{\Psi}^T)^{-1}
\mathbf{\beta}^2\mathbf{1}_L\\
\mathbf{q}&=\sigma^{2}(\mathbf{\varepsilon} - \mathbf{D} -\mathbf{\beta}^2\mathbf{\Psi})^{-1}
\mathbf{\beta}^2\mathbf{1}_L,
\end{split}
\end{equation}
where $\mathbf{1}_L$ is the length-$L$ vector consisting of all
ones,$\mathbf{\beta}~=~\mathrm{diag}\left\{\beta_1,\ldots,\beta_L\right\}$,
$\mathbf{D}$ is diagonal with
\begin{equation}\label{eqn:D}
\left[\mathbf{D}\right]_{l,l}=|\beta_l\tilde{\mathbf{h}}_l^H\bar{\mathbf{u}}_l|^2 -
2\beta_l\mathfrak{Re}\left[\tilde{\mathbf{h}}_l^H\bar{\mathbf{u}}_l\right] + 1,
\end{equation}
where $\mathfrak{Re}\left[\cdot\right]$ denotes the real part of a
complex number and
\begin{equation}\label{eqn:Psi}
[\mathbf{\Psi}]_{ij}=\left\{ \begin{array}{ll}
|\mathbf{\tilde{h}}_{i}^{H}\bar{\mathbf{u}}_{j}|^{2} & \textrm{${i}\neq{j}$}\\
0 & \textrm{$i=j$}\end{array} \right..
\end{equation}

\section{Equality of Downlink and Uplink Power Allocations\label{section:proof}}
Based on~(\ref{eqn:poweralloc}), we see that
$\mathbf{\Psi}=\mathbf{\Psi}^T$ is a sufficient condition for the
equality of $\mathbf{p}$ and $\mathbf{q}$.  We now proceed to prove
that this transpose symmetry indeed applies for arbitrary virtual
uplink precoders $\bar{\mathbf{V}}_k$ as long as the optimum power
allocation $\mathbf{q}^\star=\left[q^\star_1,\dots,q^\star_L\right]$
satisfying~(\ref{eqn:optPowerAlloc}) and the corresponding MMSE receive
beamformers $\mathbf{u}^\star_l$ are used.

\subsection{KKT Conditions for MMSE Precoding}
From the objective and constraint functions
in~(\ref{eqn:optPowerAlloc}), the Lagrangian is
\begin{equation}\label{eqn:lagrangian}
\begin{split}
\mathcal{L}\left(\mathbf{q},\mathbf{\mu}\right) = & \mathrm{tr}\left[\left(\sum_{l=1}^L q_l
\tilde{\mathbf{h}}_l\tilde{\mathbf{h}}_l^H + \sigma^2\mathbf{I}_M\right)^{-1}\right]\\
& + \mu_{\mathrm{sum}}\left(\sum_{l=1}^L q_l - P_{\max}\right) - \sum_{l=1}^L \mu_l q_l,
\end{split}
\end{equation}
and the resulting KKT conditions are
\begin{equation}\label{eqn:KKT}
\begin{split}
\nabla\mathcal{L}= &-\left[\begin{array}{c}\tilde{\mathbf{h}}_1^H\mathbf{J}^{-2}
\tilde{\mathbf{h}}_1\\\\\vdots\\\tilde{\mathbf{h}}_L^H\mathbf{J}^{-2}\tilde{\mathbf{h}}_L
\end{array}\right] + \mu_{\mathrm{sum}}\mathbf{1}_L - \sum_{l=1}^L \mu_l
\mathbf{e}_l=\mathbf{0}_L \\
&\quad \sum_{l=1}^L q_l \le P_{\max}, \quad q_l \ge 0\\
&\quad \mu_{\mathrm{sum}} \ge 0, \mu_l \ge 0\\
&\quad \mu_{\mathrm{sum}} \left(\sum_{l=1}^L q_l - P_{\max}\right) = 0,\quad \mu_lq_l = 0.
\end{split}
\end{equation}
Here, $\mathbf{0}_L$ is the length-$L$ all-zeroes vector, and
$\mathbf{e}_l$ is the standard basis vector with a single one in the
$l^{\mathrm{th}}$ position and zeroes elsewhere.  The gradient in the
stationarity condition follows from the identity $\partial
\mathbf{J}^{-1}
=-\mathbf{J}^{-1}\left(\partial\mathbf{J}\right)\mathbf{J}^{-1}$~\cite{MCB}
and the linearity of the trace operator. Thus,
\begin{equation}
\begin{split}
\frac{\partial \mathrm{tr}\left[\mathbf{J}^{-1}\right]}{\partial q_l} & = \mathrm{tr}
\left[-\mathbf{J}^{-1}\frac{\partial\mathbf{J}}{\partial q_l}\mathbf{J}^{-1}
\right] = -\mathrm{tr}\left[\mathbf{J}^{-1}\tilde{\mathbf{h}}_l\tilde{\mathbf{h}}_l^H
\mathbf{J}^{-1}\right]\\
& = -\tilde{\mathbf{h}}_l^H\mathbf{J}^{-2}\tilde{\mathbf{h}}_l.
\end{split}
\end{equation}

\subsection{Conditions for Equality under Optimal Power Allocation}
\label{subsec:conditions}

Having solved~(\ref{eqn:optPowerAlloc}) for an arbitrary set of virtual
uplink precoders $\bar{\mathbf{v}}_l$, we then find the MMSE receive
beamformers
$\mathbf{u}^\star_l=\mathbf{J}^{-1}\tilde{\mathbf{h}}_l\sqrt{q^\star_l}$.
With the associated virtual uplink stream MSEs $\varepsilon_l$ and scalars
$\beta_l=\sqrt{q^\star_l}\|\mathbf{u}^\star_l\|$, we can then
use~(\ref{eqn:poweralloc}) to find the downlink power allocation
$\mathbf{p}$ that achieves the same MSEs for each data stream.

In the case where the optimal power allocation results in one or more
\textit{inactive streams} $\mathcal{S}_I = \left\{l \in (1,\dots,L) \;
| \; q^\star_l=0\right\}$, this algorithm fails since
$\mathbf{u}^\star_l = \mathbf{0}$ for $l \in \mathcal{S}_I$.  However,
the same MSEs can be achieved for these inactive streams in the
downlink by setting $p_l=0$.  The power allocation $\mathbf{p}$ for the
set of \textit{active streams} $\mathcal{S}_A = \left\{l \in
(1,\dots,L) \; | \; q^\star_l>0\right\}$ can then be found by following
the specified procedure after deleting the rows and columns from
$\mathbf{\beta}$, $\mathbf{D}$, and $\mathbf{\Psi}$ corresponding to
the inactive streams.

The coupling matrix $\mathbf{\Psi}$ is a real matrix whose off-diagonal
entries $\left[\mathbf{\Psi}\right]_{ij}$ contain squared magnitudes of
the end-to-end channel gains from transmitted symbol $x_j$ to the
decoded symbol $\hat{x}_i$.  We observe that
$\mathbf{\Psi}=\mathbf{\Psi}^T$ is satisfied when
\begin{equation}
\begin{split}
\frac{\tilde{\mathbf{h}}_i^H\mathbf{u}^\star_j}{\|\mathbf{u}^\star_j\|} & =
\frac{{\mathbf{u}^\star_i}^H\tilde{\mathbf{h}}_j}{\|\mathbf{u}^\star_i\|},
\end{split}
\end{equation}
or equivalently,
\begin{equation}
\begin{split}
\frac{\tilde{\mathbf{h}}_i^H\mathbf{J}^{-1}\tilde{\mathbf{h}}_j\sqrt{q^\star_j}}
{\sqrt{q^\star_j\tilde{\mathbf{h}}_j^H\mathbf{J}^{-2}\tilde{\mathbf{h}}_j}} & =
\frac{\sqrt{q^\star_i}\tilde{\mathbf{h}}_i^H\mathbf{J}^{-1}\tilde{\mathbf{h}}_j}
{\sqrt{q^\star_i\tilde{\mathbf{h}}_i^H\mathbf{J}^{-2}\tilde{\mathbf{h}}_i}}\
\end{split}
\end{equation}
The power allocation terms $q^\star_i$ and $q^\star_j$ cancel out, and
numerators are equal; thus, an equivalent expression for the sufficient
condition for $\mathbf{p}=\mathbf{q}$ is
\begin{equation}\label{eqn:sufficient}
\tilde{\mathbf{h}}_i^H\mathbf{J}^{-2}\tilde{\mathbf{h}}_i = \tilde{\mathbf{h}}_j^H
\mathbf{J}^{-2}\tilde{\mathbf{h}}_j \quad \forall i,j \in \mathcal{S}_A.
\end{equation}

We rewrite the individual terms in~(\ref{eqn:KKT}) as
$\tilde{\mathbf{h}}_l^H\mathbf{J}^{-2}\tilde{\mathbf{h}}_l =
\left(\mu_{\mathrm{sum}}-\mu_l\right)$.  Due to the complementary
slackness condition ($\mu_l q_l=0$), the dual variables $\mu_l$ are
zero for all active streams $l \in \mathcal{S}_A$ with $q_l>0$.  Thus,
it follows that
\begin{equation}\label{eqn:sufficientProof}
\tilde{\mathbf{h}}_l^H\mathbf{J}^{-2}\tilde{\mathbf{h}}_l = \mu_{\mathrm{sum}}
\quad \forall l \in \mathcal{S}_A;
\end{equation}
that is,~(\ref{eqn:sufficient}) is satisfied,
$\mathbf{\Psi}=\mathbf{\Psi}^T$, and the downlink and virtual uplink
power allocations $\mathbf{p}$ and $\mathbf{q}$ that achieve the same
minimum sum-MSE are identical.

\subsection{Discussion}
The equality result presented in Section~\ref{subsec:conditions} was
shown to apply for arbitrary $\bar{\mathbf{v}}_l$, as long as the
optimum power allocation and MMSE receivers are used.  It follows that
it also applies to the optimum covariance-based design, when covariance
matrices for each stream are normalized as $\mathbf{R}_l =
q^\star_l\bar{\mathbf{R}}_l$ and
$\bar{\mathbf{R}}_l=\bar{\mathbf{v}}_l\bar{\mathbf{v}}_l^H$.   This result implies
that the virtual uplink to downlink transformation stage can be omitted
from algorithms using both iterative and joint designs based on a
virtual-uplink solution~\cite{SSJB05,KTA06,MJHU06,SSB07}, thus allowing
for simplified implementations.

\section{Conclusions\label{section:conclusions}}
In this letter, we have proven that the optimum power allocations for
the downlink and virtual uplink are identical when minimizing the
sum-MSE under a sum power constraint.  With this proof, we extend the
known results in a well studied problem. Importantly, our result
simplifies existing iterative algorithms, eliminating the solution of a
matrix equation in each iteration.

\bibliographystyle{IEEEtran}
\bibliography{IEEEabrv,PowerAllocLetter}
\newpage

\end{document}